\documentstyle[11pt]{article}

\begin{document}

\input{epsf}

\def\beq{\begin{equation}}
\def\eeq{\end{equation}}
\def\bea{\begin{eqnarray}}
\def\eea{\end{eqnarray}}
\def\beas{\begin{eqnarray*}}
\def\eeas{\end{eqnarray*}}
\def\ov{\overline}
\def\ot{\otimes}

\newcommand{\hf}{\mbox{$\frac{1}{2}$}}
\def\sig{\sigma}
\def\De{\Delta}
\def\af{\alpha}
\def\be{\beta}
\def\la{\lambda}
\def\ga{\gamma}
\def\ep{\epsilon}
\def\vep{\varepsilon}
\def\half{\frac{1}{2}}
\def\third{\frac{1}{3}}
\def\fth{\frac{1}{4}}
\def\sth{\frac{1}{6}}
\def\tth{\frac{1}{24}}
\def\tde{\frac{3}{2}}

\def\zb{{\bar z}} 
\def\psib{{\bar \psi}} 
\def\etab{{\bar \eta }}
\def\gab{{\bar \ga}}
\def\vev#1{\langle #1 \rangle}
\def\inv#1{{1 \over #1}}

\def\CA{{\cal A}}       \def\CB{{\cal B}}       \def\CC{{\cal C}}
\def\CD{{\cal D}}       \def\CE{{\cal E}}       \def\CF{{\cal F}}
\def\CG{{\cal G}}       \def\CH{{\cal H}}       \def\CI{{\cal J}}
\def\CJ{{\cal J}}       \def\CK{{\cal K}}       \def\CL{{\cal L}}
\def\CM{{\cal M}}       \def\CN{{\cal N}}       \def\CO{{\cal O}}
\def\CP{{\cal P}}       \def\CQ{{\cal Q}}       \def\CR{{\cal R}}
\def\CS{{\cal S}}       \def\CT{{\cal T}}       \def\CU{{\cal U}}
\def\CV{{\cal V}}       \def\CW{{\cal W}}       \def\CX{{\cal X}}
\def\CY{{\cal Y}}       \def\CZ{{\cal Z}}

\newcommand{\np}{Nucl. Phys.}
\newcommand{\pl}{Phys. Lett.}
\newcommand{\prl}{Phys. Rev. Lett.}
\newcommand{\cmp}{Commun. Math. Phys.}
\newcommand{\jmp}{J. Math. Phys.}
\newcommand{\jpamg}{J. Phys. {\bf A}: Math. Gen.}
\newcommand{\lmp}{Lett. Math. Phys.}
\newcommand{\ptp}{Prog. Theor. Phys.}

\newif\ifbbB\bbBfalse                
\bbBtrue                             

\ifbbB   
 \message{If you do not have msbm (blackboard bold) fonts,}
 \message{change the option at the top of the text file.}
 \font\blackboard=msbm10 
 \font\blackboards=msbm7 \font\blackboardss=msbm5
 \newfam\black \textfont\black=\blackboard
 \scriptfont\black=\blackboards \scriptscriptfont\black=\blackboardss
 \def\Bbb#1{{\fam\black\relax#1}}
\else
 \def\Bbb{\bf}
\fi

\def\id{{1\! \! 1 }}
\def\bo{{\Bbb 1}}
\def\bI{{\Bbb I}}
\def\bC{{\Bbb C}} 
\def\bZ{{\Bbb Z}}
\def\CN{{\cal N}}

\title{Non-additive fusion, Hubbard models and non-locality}
\author{{\bf Z. Maassarani}\thanks{Work supported  by DOE grant no.
DE-FG02-97ER41027 and NSF grant no. DMR-9802813.} \\
\\
{\small Physics Department}\\
{\small University of Virginia}\\
{\small 382 McCormick Rd.}\\
{\small Charlottesville, VA,  
22903 USA}\thanks{Email address: zm4v@virginia.edu} \\}
\date{}
\maketitle

\begin{abstract}
In the framework of quantum groups and additive $R$-matrices, 
the fusion procedure allows to construct higher-dimensional solutions of the 
Yang-Baxter equation. These solutions lead to integrable one-dimensional 
spin-chain Hamiltonians.  Here fusion is shown to generalize naturally 
to non-additive $R$-matrices, which therefore do not 
have a quantum group symmetry.
This method is then applied  to the generalized Hubbard models.
Although the resulting integrable models are
not as simple as the starting ones,  the general structure is 
that of two  spin-$(s\times s')$  $sl(2)$ models coupled at 
the free-fermion point. An important issue is the probable
lack of  regular points which give local Hamiltonians.
This problem is related to the existence of 
second order zeroes in the unitarity equation, 
and  arises   for the   XX models of higher spins,
the building blocks of the Hubbard models.
A possible  connection
between some  Lax operators $L$ and  $R$-matrices is noted. 
\end{abstract}
\vspace*{2.5cm}
\noindent
\hspace{1cm} July 1999\hfill\\

\thispagestyle{empty}

\newpage

\setcounter{page}{1}

\section{Introduction}

The construction and diagonalization of integrable one-dimensional spin-chain
Hamiltonians within the framework of the Quantum Inverse Scattering Method is 
well-known \cite{qism1,qism2,qism3,kbi}. A given integrable model and
all its conserved quantities are encoded in an $R$-matrix which satisfies 
the Yang-Baxter equation. The quantum group approach \cite{jimbo1,drinfeld} 
provides a systematic   way  for obtaining a large class of 
solutions based on representations
of some underlying Lie algebra \cite{jimbo2} or super-algebra \cite{bgz}.
By construction, such solutions possess the additivity property which means
that the initial double spectral parameter dependence reduces to the  
difference of the two parameters. The most famous example is the spin-$\half$ 
XXZ chain. 

On the other hand models having non-additive $R$-matrices have been known
to exist for a long time. Examples of such models include Shastry's 
solution for the Hubbard model \cite{sh12,sh3}, the Chiral Potts models
\cite{bapo,alls,ampty} and more recently the Bariev 
models \cite{bariev,hqz,sw}.
The Hubbard model both in its bosonic and fermionic guises  
was also generalized to multi-state versions
while retaining  the same algebraic structure. Initial versions were
first introduced in  \cite{zm1}, studied in \cite{zm2,martins,ys},
further generalized in \cite{ff} and fermionized in \cite{pz}. 
(See also \cite{gomu} for another possible fermionization scheme.)
All these non-additive solutions of the Yang-Baxter equations are 
isolated and do not yet fit in a general framework. 

A  general method for constructing solutions to the Yang-Baxter 
equation out of a  given known one is the fusion method. It works by
multiplying the same matrix by itself a certain number
of times, at different values of the spectral parameter,  
and finally  multiplying  by a projector. This works much the same
way as building higher-dimensional representations from 
tensor products  of a  smaller one and a final projection on a subspace. 
For instance an $sl(2)$ spin-$s$ solution
can be obtained by successive fusions of the spin-$\half$ 
solution \cite{krs}. 
It is in fact possible to fuse an arbitrary product of  $R$ matrices to obtain solutions of the YBE corresponding to most (but not always all \cite{zmosp})
representations of a given  Lie algebra  \cite{kr,kns}. 
In the framework of quantum groups,
the direct method for finding $R$-matrices  with a given  Lie algebra symmetry,
and corresponding to a given representation, 
consists of solving linear equations
\cite{jimbo2}. This method and fusion give the same results. 

In \cite{zmfuse} fusion was shown to work for  a class of models
which retained only   some aspects of an $sl(m)$ 
quantum group structure.  Higher-dimensional solutions were obtained by fusion,
where  no quantum group symmetry and therefore no  direct method existed. 
 
In this work I derive  fusion equations for non-additive $R$-matrices.
The results of  section \ref{nonaddfus}
are quite general and require the starting matrix to satisfy 
only a minimal number of properties. 
The generalized Hubbard models are shown to satisfy these
properties. This allows to construct 
higher-spin Hubbard models which appear as two copies of a multi-flavor
spin-$(s\times s')$ model  coupled at the `free-fermion' point.
The coupling does not have the simple structure of the starting models. 
The resulting integrable models have non-additive, unitary,  
$R$-matrices but appear to lack the usual regularity property, 
which would allow to obtain {\it local} Hamiltonians. 
The source of this lack of regularity is traced back to
the  spin-$s$  building blocks. These models, for $s\geq 1$,
are not regular but still allow for local mutually commuting 
quantities through a limiting procedure from a generic $q$ value. 
It is not clear how to implement this for the higher-spin 
Hubbard models. Possible applications to the Bariev and Chiral Potts models 
are mentioned in the conclusion. A possible connection
between some Lax matrices and $R$-matrices is also proposed.

\section{Fusion for non-additive $R$-matrices}\label{nonaddfus}

Given a non-additive solution of the Yang-Baxter equation, 
it is possible to obtain new solutions provided one has a projector point.
Expressions for four fused matrices are found along with the equations they 
satisfy. The results of this section are general and hold 
without  reference to any 
particular model. The notation follows closely that of reference \cite{zmfuse}.
The word fusion here is used in the conventional sense, and not in the sense
of \cite{ff}.
For the additive case, Kulish and Sklyanin had already realized
that only two properties were needed for fusion to be possible.
The Yang-Baxter equation has to be satisfied and a projector point must 
exist (p. 108 of \cite{qism1}). 

Consider a non-additive solution $R(\la_1,\la_2)$ of the Yang-Baxter 
equation (YBE)
\beq
R_{12}(\la_1,\la_2) R_{13}(\la_1,\la_3)
R_{23}(\la_2,\la_3) = R_{23}(\la_2,\la_3) 
R_{13}(\la_1,\la_3) R_{12}(\la_1,\la_2)\label{rybe}
\eeq
Additivity  means that for a proper choice of parameterization, and after eventual transformations such as a
gauge (a special similarity transformation on $R$) \cite{jimbo2}  
or twist (a special similarity transformation on $\check{R}$)
\cite{twist} transformation, one can write
$R_{12}(\la_1,\la_2)=R_{12}(\la_1-\la_2)$. Most known solutions of the 
YBE are additive. This includes in particular all  solutions corresponding 
to a quantum group symmetry, $\CU_q(\CG)$, where $\CG$ is any Lie algebra
or super-algebra \cite{jimbo2,bgz}.
Known non-additive solutions include the class of  generalized Hubbard
models in  their bosonic and fermionic forms \cite{ff,pz}, the chiral
Potts models \cite{ampty} and the Bariev models \cite{bariev,hqz,sw}.

To implement fusion it is enough that the solution
at hand has a projector point. Thus
consider  any solution $R$ of the Yang-Baxter equation (\ref{rybe}),
which becomes proportional to  a projector at some special values of 
the spectral parameters pair $(\la_1,\la_2)$. 
Specifically, define the projector $\pi^{(1)}$ through
\beq
R(\la+\rho,\la)\equiv g(\la)\, \pi^{(1)}\label{proj}
\eeq
and let $\pi^{(2)}=\bI-\pi^{(1)}$
be the orthogonal complementary projector. 
Both projectors are assumed to be independent of $\la$, and $\rho$ is some
fixed value characteristic of the $R$-matrix at hand. 
I have  then  verified that the methods used in \cite{zmfuse} can 
be extended  to the non-additive setting.
This yields the following results.  

Let $S$ be the matrix which diagonalizes both projectors. 
Define two  fused matrices, for $i=1,2$,  by 
\beq
R^{(i)}_{<12>3}(\la,\la_3) = S^{-1}_{12}
\pi^{(i)}_{12}\, R_{13}(\la,\la_3)\, R_{23}(\la-\rho,\la_3) \,\pi^{(i)}_{12}S_{12}\label{fus1}
\eeq
The matrices (\ref{fus1})  satisfy a YBE where one space is a tensor
product of two spaces:
\bea 
& & R^{(i)}_{<12>3}(\la,\la_3) \, R^{(i)}_{<12>4}(\la,\la_4) \, 
R_{34}(\la_3,\la_4) \nonumber\\
& & = R_{34}(\la_3,\la_4) \,R^{(i)}_{<12>4}(\la,\la_4)\,
R^{(i)}_{<12>3}(\la,\la_3) 
\;\;,\;\;\; i=1,2 \label{ybefus1}
\eea
We  have thus obtained  new $R$-matrices. 
If $d_i, i=1,2,3$, are the dimensions of the spaces 1, 2 and 3,
and ${\rm tr}(\pi^{(1)})= d$, 
then  after deletion of the vanishing rows and columns, 
$R^{(i)}_{<12>3}(\la,\la_3)$ is a $d d_3$ dimensional matrix for $i=1$,
and  $(d_1 d_2 -d) d_3$ dimensional for $i=2$. 

Note also that  there is another possible  choice
of fused matrices obtained by taking the right-hand side 
of the YBE at the projector point (see (4) of \cite{zmfuse}).
There is however no essential difference with the foregoing choice. 

One can then fuse two matrices $R_{<12>3}^{(i)}(\la,\la_3)$
to obtain the  matrix $R_{<12><34>}^{(i)}(\la,\mu)$ defined by: 
\beq
R^{(i)}_{<12><34>}(\la,\mu) =
S^{-1}_{34}\pi^{(i)}_{34}\, R_{<12>4}^{(i)}(\la,\mu-\rho)\, 
R_{<12>3}^{(i)}(\la,\mu)
\,\pi^{(i)}_{34}S_{34}\;\;,\;\;\; i=1,2\label{fus3}\\
\eeq
These matrices have dimensions $d^2$  for $i=1$,
and  $(d_1 d_2 -d)^2$  for $i=2$.
They satisfy two Yang-Baxter equations ($i=1,2$):
\bea
& & R^{(i)}_{<12><34>}(\la,\mu) \, R^{(i)}_{<12>5}(\la,\la_5) \, 
R_{<34>5}^{(i)}(\mu,\la_5) \nonumber\\
& & = R_{<34>5}^{(i)}(\mu,\la_5)\, R^{(i)}_{<12>5}(\la,\la_5)\,  
R^{(i)}_{<12><34>}(\la,\mu) \label{ybefus2}\\
& & R^{(i)}_{<12><34>}(\la,\mu) \, R^{(i)}_{<12><56>}(\la,\nu) \, 
R_{<34><56>}^{(i)}(\mu,\nu)    \nonumber\\
& & =  R_{<34><56>}^{(i)}(\mu,\nu)\, R^{(i)}_{<12><56>}(\la,\nu)\,  
R^{(i)}_{<12><34>}(\la,\mu)  \label{ybefus3}
\eea

Assume  now  that the original $R$-matrix is  regular and unitary, 
{\it i.e.}
\bea
& & R_{12}(\mu,\mu)= c(\mu)\, \CP_{12}\label{regun1}\\ 
& & R_{12}(\la,\mu)\, R_{21}(\mu,\la)= f(\la,\mu)\, \bI\label{regun2}
\eea
where $\CP$ is the permutation operator and  
$R_{21}\equiv\CP_{12} R_{12} \CP_{12}$. 
The function $f(\la,\mu)$ is then symmetric in its arguments, 
and $c(\mu)$ is some complex, generically non-vanishing function.
The fused matrices (\ref{fus3}) inherit the regularity property (\ref{regun1})
provided they are correctly normalized, for the corresponding $c$-function  
not to vanish. This can be  achieved  by the following  normalization. 
Insert a factor of  $\left( f(\la+\rho,\mu) \right)^{-1}$ in  the 
right-hand side of equation (\ref{fus3}). 
As $R(\mu+\rho,\mu)$ is a non-trivial  projector, unitarity (\ref{regun2})
implies that $f(\mu+\rho,\mu)$ vanishes for all values of $\mu$.
The normalization just introduced  cancels this zero in the numerator and
leaves a regular fused-matrix. Incidentally, the symmetry of $f$ implies
that  $f(\mu,\mu+\rho)$ also vanishes. 

The form (\ref{fus3}) can be simplified to a more symmetric one:
\bea
R^{(i)}_{<12><34>}(\la,\mu) &=&\frac{1}{f(\la+\rho,\mu)}
S^{-1}_{12}S^{-1}_{34}\pi^{(i)}_{12}\pi^{(i)}_{34}\,
R_{14}(\la,\mu-\rho)\,  R_{24}(\la-\rho,\mu-\rho)\nonumber\\
&\times&  R_{13}(\la,\mu)\, R_{23}(\la-\rho,\mu)\,\pi^{(i)}_{12}
\pi^{(i)}_{34}S_{12}S_{34}\;\;,\;\;\; i=1,2\label{fus4}
\eea
where the normalization has been included. 

Let $\partial_i f(\la_1,\la_2)$ denote the derivative with respect to 
the $i^{\rm th}$ slot ($i=1,2$). 
Taking the limit $\la\longrightarrow \mu$ for the 
normalized matrices I find:
\bea
R^{(1)}_{<12><34>}(\mu,\mu) &=& c(\mu)c(\mu-\rho)\,
\frac{\partial_2 f(\mu,\mu-\rho)}{\partial_1 
f(\mu+\rho,\mu)}\, \CP_{13} \CP_{24}\, 
S_{12}^{-1}\pi^{(1)}_{12}
S_{12}\, S_{34}^{-1} \pi^{(1)}_{34} S_{34}\label{regs1}\\ 
R^{(2)}_{<12><34>}(\mu,\mu) &=&  c(\mu)c(\mu-\rho)\, 
\frac{\partial_1 f(\mu,\mu-\rho)}{\partial_1 f(\mu+\rho,\mu)} 
\,\CP_{13} \CP_{24}\, 
S_{12}^{-1}\pi^{(2)}_{12}
S_{12}\, S_{34}^{-1} \pi^{(2)}_{34} S_{34}\label{regs2}
\eea
The unitarity property is  inherited independently from the
normalization:
\bea
R^{(i)}_{<12><34>}(\la,\mu)  R^{(i)}_{<34><12>}(\mu,\la) &=& 
\frac{f(\la-\rho,\mu) f(\la,\mu)f(\la-\rho,\mu-\rho)
f(\la,\mu-\rho)}{f(\la+\rho,\mu)f(\mu+\rho,\la)}\nonumber\\
& & \times S_{12}^{-1} \pi^{(i)}_{12} S_{12}\,  
S_{34}^{-1} \pi^{(i)}_{34} S_{34}\;\;,\;\;\; i=1,2\label{units}
\eea
where $R^{(i)}_{<34><12>}(\la,\mu)= \CP_{13} \CP_{24}\, 
R^{(i)}_{<12><34>}(\la,\mu)\, \CP_{13}\CP_{24}$.

In the framework of the QISM, the 
quadratic Hamiltonian density of such integrable hierarchies is 
the derivative at $\la=\mu$ of the matrix 
$\check{R}(\la,\mu)=\CP R(\la,\mu)$. 
Taking the limit yields:
\bea
\frac{d}{d\la}\check{R}^{(i)}_{<12><34>}(\la,\mu)_{|_{\la=\mu}} &=&
- \frac{\partial_1^2 f(\mu+\rho,\mu)}{2 \partial_1 f(\mu+\rho,\mu)} \, 
\check{R}^{(i)}_{<12><34>}(\mu,\mu) \nonumber\\
& &+ \frac{1}{2 \partial_1 f(\mu+\rho,\mu)} \, 
S_{12}^{-1} S_{34}^{-1} \pi^{(i)}_{12} \pi^{(i)}_{34}
\,\frac{d^2}{d\la^2}\left( R_{32}(\la,\mu-\rho) \check{R}_{13}(\la,\mu)
\right. \nonumber\\
& & \left.\times \check{R}_{24}(\la-\rho,\mu-\rho) R_{23}(\la-\rho,\mu)\right)_{|_{\la=\mu}}\,
\pi^{(i)}_{12} \pi^{(i)}_{34} S_{12} S_{34} \label{hamden}
\eea 
The first term is proportional to the identity in the fused
spaces and may be dropped.
The results of \cite{zmfuse} in the additive case can be recovered
by setting $f(\la,\mu)\longrightarrow f(\la-\mu)$, with $f$ now an even
function. (There is an obvious misprint in 
formula (17) of \cite{zmfuse}: $f'(0)$ and $f''(0)$ should be 
replaced by $f'(\rho)$ and $f''(\rho)$.)

An implicit assumption, which does {\it not} affect the Yang-Baxter equations,  
was made when deriving the regularity 
equations (\ref{regs1}) and (\ref{regs2}): $f(\la+\rho,\mu)$ was
taken to vanish like $(\la-\mu)$ for $\la\longrightarrow \mu$. 
However this zero can be of second order. This will be case for 
the Hubbard models studied in the next section. A zero of any order
does not affect the unitarity equation (\ref{units}) as numerator and 
denominator compensate to give a finite non-vanishing result. But
it is necessary to do a second or higher 
order expansion to find the appropriate
expressions for the regularity equations. It does not appear possible 
to prove in general, and with a minimal set of assumptions, 
that regularity still holds. But the result is finite. An argument in favor
of regularity is  the unitarity equation which says $\check{R}_{12}(\mu,\mu)$ 
squares to the identity. However this turns out not to be enough and
specific counter-examples are the multi-flavor spin-$s$  
models at $\gamma=\pi/2$, for $s\geq 1$. 
Note that the above issues of regularity will arise for all  fused
matrices with non-simple zeroes in the unitarity equations of  the
starting $R$-matrix. 

In the case of a second order   
zero for $f(\la,\mu)$, and provided the fused matrix is regular,
the  Hamiltonian density is given by:
\bea
\frac{d}{d\la}\check{R}^{(i)}_{<12><34>}(\la,\mu)_{|_{\la=\mu}} &=&
- \frac{\partial_1^3 f(\mu+\rho,\mu)}{3 \partial_1^2 f(\mu+\rho,\mu)} \, 
\check{R}^{(i)}_{<12><34>}(\mu,\mu) \nonumber\\
& &+ \frac{1}{3 \partial_1^2 f(\mu+\rho,\mu)}  
S_{12}^{-1} S_{34}^{-1} \pi^{(i)}_{12} \pi^{(i)}_{34}
\,\frac{d^3}{d\la^3}\left( R_{32}(\la,\mu-\rho) \check{R}_{13}(\la,\mu)
\right. \nonumber\\
& & \left.\times \check{R}_{24}(\la-\rho,\mu-\rho) R_{23}(\la-\rho,\mu)\right)_{|_{\la=\mu}}\,
\pi^{(i)}_{12} \pi^{(i)}_{34} S_{12} S_{34} \label{hamden2}
\eea
Regularity implies that the first term is proportional to the identity and may 
be dropped. 

The projector property is the only additional ingredient
needed to construct fused matrices.
For the $R$-matrices based on Lie algebras, 
the degeneration of a generically invertible  $R$-matrix  
to a projector is expected.  
But for non-additive matrices one has to verify in every case 
whether a projector point exists. 
Note also that the unitarity property (\ref{units})  for the fused
matrix indicates that it may have its own projector point. This 
in turn implies that fusion may be continued to another level, or
even indefinitely as happens in the quantum group framework. 

The above fusing scheme will now be applied to  the 
generalized Hubbard models.

\section{Hubbard fusion and  non-locality}\label{hubfus}

We first recall  the construction of the multi-state or multi-flavor Hubbard
models in their bosonic form \cite{ff}. The connection
between the $L$ and $R$ matrices is clarified.  Fusion is  implemented. 
The connection between double zeroes in the unitarity equation and
the lack of regularity is discussed on specific examples. 

\subsection{A Hubbard primer}\label{hubpri}

The following `free-fermions' or XX models are building blocks of 
the Hubbard models. 
Let $n$, $n_1$ and $n_2$ be three positive integers such that $n_1+n_2=n$, 
and $A$, $B$ be two disjoint sets whose union is the set of basis 
states of $\bC^n$, with card$(A)=n_1$ and card$(B)=n_2$. 
Let $E^{\af\be}$ be a square matrix  with a one at row $\af$ 
and column $\be$ and zeroes otherwise.
Define
\bea
\tilde{P}^{(1)}&=&\sum_{a\in A}\sum_{\beta\in B}\left(E^{a\beta}\otimes
E^{\beta a} + E^{\beta a}\otimes E^{a\beta}\right)\\
\tilde{P}^{(2)}&=&\sum_{a,a'\in A} E^{a a'}\otimes
E^{a' a} + \sum_{\beta,\beta'\in B}
E^{\beta\beta'}\otimes E^{\beta' \beta}\\
\tilde{P}^{(3)}&=&\sum_{a\in A}\sum_{\beta\in B}\left(x E^{a a}\otimes
E^{\beta\beta} + x^{-1} E^{\beta\beta}\otimes E^{a a}\right)
\eea
Latin indices always belong to $A$ while Greek indices belong to $B$.
The complex twist parameter $x$ is arbitrary. 
The free-fermions $R$-matrix
\beq
R(\la)= \tilde{P}^{(1)} +\tilde{P}^{(2)} \cos\la + \tilde{P}^{(3)} \sin\la
\label{ffr}
\eeq
satisfies the additive Yang-Baxter equation:
\beq
R_{12}(\la-\mu) \, R_{13}(\la) \, R_{23}(\mu) =
R_{23}(\mu) \, R_{13}(\la) \, R_{12}(\la-\mu) \label{ybeadd}
\eeq
The interpretation of the multiple-flavors  in terms of $sl(2)$ states was  
done in \cite{zmfuse}.

Coupling two commuting copies of the foregoing models gives 
the Hubbard models. Where made explicit,
the two copies are denoted by unprimed and primed quantities.
Let us stress that the
copies need {\it not} be of the same type. For instance, the `left'
copy can be  $(n_1,n_2)$ while the `right' copy is $(n'_1,n'_2)$
with $n$ not necessarily equal to $n'$. 
($n_i=n'_i=1$
correspond to the original Hubbard model.)
The twist parameters may also differ. 
One then  defines a multi-flavor version of $\sigma^z$, 
the conjugation matrix
\beq
C=\sum_{\beta\in B} E^{\beta\beta} -\sum_{a\in A} E^{aa}\label{cc}
\eeq
and a diagonal coupling matrix
\beq
I_{00'}(h)  =\cosh \left(\frac{h}{2}\right)\, \bI + 
\sinh \left(\frac{h}{2}\right) \; C_0 \,C'_0 
=\exp\left(\frac{h}{2}\, C_0 \,C'_0\right)\label{imat}
\eeq
The parameter $h$ is related to the spectral parameter $\lambda$ by 
\beq
\sinh (2h) =  U \sin (2\la) \label{rela}
\eeq
where $U$ is the coupling constant. 
One chooses for $h(\la)$ the principal branch which vanishes for 
vanishing $\la$ or $U$.  
The Lax operator at site $i$ is equal to:
\beq
L_{0i}(\la) = I_{00'}(h)\, R_{0i}(\la)\, R_{0'i'}(\la)\, I_{00'}(h)
\eeq
Their commutation relations for different spectral parameters 
at a given site are given by
\beq
R(\la_1,\la_2) \; \stackrel{1}{L}(\la_1) \stackrel{2}{L}(\la_2) =  
\stackrel{2}{L}(\la_2) \stackrel{1}{L}(\la_1)
\;R(\la_1,\la_2)\label{rll}
\eeq 
where $\stackrel{1}{L}(\la_1)=L(\la_1)\otimes \bI$, 
$\stackrel{2}{L}(\la_2)=\bI\otimes L(\la_2)$, and 
\bea
R(\la_1,\la_2)&=&  I_{12}(h_1) I_{34}(h_2) \left[
R_{13}(\la_1-\la_2)  R_{24}(\la_1-\la_2)
+\frac{\sin(\la_1-\la_2)}{\sin(\la_1+\la_2)}\right.\nonumber\\
& \times&\left. \tanh(h_1+h_2) 
R_{13}(\la_1+\la_2) C_1 R_{24}(\la_1+\la_2)C_2 
\,\right]\, I_{12}(-h_1) I_{34}(-h_2)\label{rh}
\eea
This matrix is non-additive as it is not possible 
to reduce is spectral parameter dependence to $\la_1-\la_2$. 
It  satisfies the regularity property 
\beq
R(\la_1,\la_1) =  \CP_{13} \CP_{24}\label{hreg}
\eeq
and the unitarity property:
\bea
R_{12}(\la_1,\la_2) R_{21}(\la_2,\la_1)& =& 
\cos^2(\la_1-\la_2)\nonumber\\
&\times&\left(\cos^2(\la_1-\la_2) -
\cos^2(\la_1+\la_2) \tanh^2(h_1-h_2)\right) \bI \label{hunit}
\eea
The matrix  (\ref{rh}) satisfies the Yang-Baxter equation (\ref{rybe}),
with  $\la_i$ and $h_i$  related through (\ref{rela}).

The Hubbard models have a Lax matrix $L$ which differs from 
the intertwiner $R$.  
One may  wonder which matrix should
be a candidate for fusing, and what is the role of the $RLL$ relation as
opposed to the $RRR$ one.  
The matrix $R$  satisfies the symmetric  (\ref{rybe}) while $L$
satisfies the asymmetric (\ref{rll}).
This already singles out the former as the natural object to fuse. 
Another compelling reason is that $L$ and (\ref{rll}) are just
special asymmetrical limits  of $R$ and (\ref{rybe}).
Indeed one easily finds that:
\beq
R_{1234}(\la_1,0)= \frac{1}{\cosh h_1}\, I_{12}(h_1) R_{13}(\la_1)
R_{24}(\la_1)I_{12}(h_1) = \frac{1}{\cosh h_1}\, L_{(12)(34)}(\la_1)
\label{rtol}
\eeq
Setting $\la_3=0$ in (\ref{rybe}) gives (\ref{rll}). 
One also has 
\beq
R_{1234}(0,\la_2)= \frac{1}{\cosh h_2}\, I_{34}(-h_2) R_{13}(-\la_2)
R_{24}(-\la_2)I_{34}(-h_2) 
\eeq
where now the coupling of the two copies is made on the quantum spaces
rather than the auxiliary spaces. Setting $\la_1 =0$ in (\ref{rybe}) 
gives an $RLL$ relation. The corresponding quadratic Hamiltonian
and all the other conserved quantities are however essentially the same
as for the auxiliary-space coupling case. 
Now the $R$ matrix can be seen as an $L$ matrix with couplings on both
auxiliary and quantum spaces. The price of this symmetrization
is the linear combination in (\ref{rh}) and the loss of additivity. 
The quadratic Hamiltonian density obtained
from $R$ at the arbitrary regular point $\la=\mu$ is given by:
\bea
\frac{d}{d\la}\check{R}(\la,\mu)_{|\la=\mu}&=&
\frac{U\cos 2\mu}{2\cosh 2 h}
(C_3 C_4 -C_1 C_2) \\
&+&  \CP_{13}\tilde{P}_{13}^{(3)}(\cosh^2 h -C_2 C_4\sinh^2 h ) \nonumber\\
&+&  \CP_{24}\tilde{P}_{24}^{(3)}(\cosh^2 h -C_1 C_3\sinh^2 h ) \nonumber \\
&-&\half\sinh(2h) \left( p_{13}^{(3)}(C_2+C_4)+p_{24}^{(3)}(C_1+C_3)
\right)\nonumber\\
&+&\frac{U}{\cosh 2 h} \left( -\sin(2\mu)\sinh(2h)\,
(\CP_{13}\tilde{P}_{13}^{(1)}\CP_{24}\tilde{P}_{24}^{(3)}
+\CP_{13}\tilde{P}_{13}^{(3)}\CP_{24}\tilde{P}_{24}^{(1)})\right.\nonumber\\
&+& 2\sin2\mu\sinh^2 h \, (p_{13}^{(1)}p_{24}^{(3)}+p_{13}^{(3)}p_{24}^{(1)})\nonumber\\
&+&\left. (p_{13}^{(1)}+\cos(2\mu) p_{13}^{(2)}+ \sin(2\mu) p_{13}^{(3)})\,
(p_{24}^{(1)}+\cos(2\mu) p_{24}^{(2)}+ \sin(2\mu) p_{24}^{(3)})
\right)\nonumber
\eea
where $h=h(\mu)$ is given by (\ref{rela}) and 
$p^{(i)}_{jk}\equiv \CP_{jk}\tilde{P}^{(i)}_{jk} C_j$, 
$i=1,2,3$ and $j,k=1,\cdots,4$.
The indices are interpreted as follows: $1\rightarrow$ site-$m$-unprimed-copy,
$2\rightarrow$ site-$m$-primed-copy, $3\rightarrow$ site-$(m+1)$-unprimed-copy,
$4\rightarrow$ site-$(m+1)$-primed-copy. 
It is only at $\mu=0$ where this expression reduces to the familiar 
generalized (bosonic) form of the Hubbard Hamiltonians.

The $I$ factors in (\ref{rh}) combine into a  similarity 
transformation. It is in fact a special type of gauge transformation,
and an equivalent $R$-matrix is given by \cite{sh3}:
\bea
r(\la_1,\la_2)&=&
R_{13}(\la_1-\la_2)  R_{24}(\la_1-\la_2)\nonumber\\
&+&\frac{\sin(\la_1-\la_2)}{\sin(\la_1+\la_2)}\tanh(h_1+h_2)\, 
R_{13}(\la_1+\la_2) C_1 R_{24}(\la_1+\la_2)C_2 \label{rrh}
\eea
The corresponding Lax matrix $l(\la)$ is given by:
\beq
r(\la,0)=\frac{1}{\cosh h}\, l(\la)= \frac{1}{\cosh h}\, 
R_{13}(\la)R_{24}(\la) I_{12}(2h)
\eeq
Similarly one finds
\beq
r(0,\la)= \frac{1}{\cosh h}\, I_{34}(-2h)R_{13}(-\la)R_{24}(-\la) 
\eeq
with a coupling on the quantum spaces. 
The regularity and unitarity properties are
satisfied without modifications. The   equivalent (for
periodic boundary conditions) Hamiltonian is simpler:
\bea
\frac{d}{d\la}\check{r}(\la,\mu)_{|\la=\mu}&=&
\CP_{13}\tilde{P}^{(3)}_{13}+\CP_{24}\tilde{P}^{(3)}_{24}+\frac{U}{\cosh 2 h}\\
&\times& (p_{13}^{(1)}+\cos(2\mu) 
p_{13}^{(2)}+ \sin(2\mu) p_{13}^{(3)})\,
(p_{24}^{(1)}+\cos(2\mu) p_{24}^{(2)}+ \sin(2\mu) p_{24}^{(3)})
\nonumber
\eea
The matrix $r$ shows that the Hubbard structure lies in the 
linear combination of two objects rather than the factors of $I(h)$.

\subsection{Fusion}

The right-hand side of   (\ref{hunit}) shows that $\la_1-\la_2=\pm \pi/2$
are possible projector points. 
Both values actually yield projectors with the same dimensionality.
This result is peculiar to the underlying XX system, but
there is otherwise no essential difference between the two projectors. 
For definiteness  $\rho =+ \pi/2$ is considered below. 
Let  
\bea
\pi^{(1)}_{1234}&=&\frac{1}{(x+x^{-1})(x^{'}+(x^{'})^{-1})} (\tilde{P}^{(1)}_{13}+\tilde{P}^{(3)}_{13})(\tilde{P'}^{(1)}_{24}+
\tilde{P'}^{(3)}_{24}) \label{proj1}\\
\pi^{(2)}_{1234}&=&\bI-\pi^{(1)}_{1234} \label{proj2}
\eea
The function $g(\la)$ defined in (\ref{proj}) is constant and equal to
$(x+x^{-1})(x'+(x')^{-1})$. The expression (\ref{proj1})
is an decoupled product of  a projector for each copy of 
a free-fermion system. To arrive at this result one uses
the following relation which is proven by a direct calculation:
\beq
{[ \pi^{(1)}_{13}\pi^{(1)}_{24},I_{12}(h)I_{34}(-h) ]}=0 \;\; , \;\;\;
\forall h\in \bC \label{commut}
\eeq
where $\pi^{(1)}_{ij}=\frac{1}{(x+x^{-1})}(\tilde{P}^{(1)}_{ij}+
\tilde{P}^{(3)}_{ij})$ is 
the projector of one copy ($x$ may have a different
value for each copy).  

The dimensions of 
these projectors are given by their traces. In particular for the 
unprimed copy, one has 
${\rm tr}\,(\pi^{(1)}) = n_1  n_2$ and  
${\rm tr}\,(\pi^{(2)}) = n_1^2 +n_2^2 + n_1 n_2$.
The matrices  which diagonalize one copy of  both projectors are given by:
\bea
S&=& \sum_{a,a'}
E^{a a}\otimes E^{a'a'}+\sum_{\beta,\beta'}
E^{\beta \beta}\otimes E^{\beta'\beta'}\nonumber\\
& & +\sum_{a}
\sum_{\beta} \left( E^{a a}\otimes E^{\beta\beta}
+x^{-1} E^{\beta\beta}\otimes E^{a a}\right)\nonumber\\
& &+\sum_{a}
\sum_{\beta}\left( E^{a\beta}\otimes E^{\beta a}
-x E^{\beta a}\otimes E^{a \beta}\right)\label{ss}\\
S^{-1}&=&\sum_{a,a'}
E^{a a}\otimes E^{a'a'}+\sum_{\beta,\beta'}
E^{\beta \beta}\otimes E^{\beta'\beta'}\nonumber\\
& & +\frac{1}{x+x^{-1}}\sum_{a}
\sum_{\beta} \left(x^{-1}E^{a a}\otimes E^{\beta\beta}
+E^{\beta\beta}\otimes E^{a a}\right)\nonumber\\
& & +\frac{1}{x+x^{-1}}\sum_{a}
\sum_{\beta}\left(-E^{a\beta}\otimes E^{\beta a}
+x E^{\beta a}\otimes E^{a \beta}\right)\label{ss1}
\eea
The diagonalized projectors  read
\bea
S^{-1}\, \pi^{(1)}\, S &=& \sum_{a}
\sum_{\beta}  E^{\beta\beta}\otimes E^{a a}\label{diagp1}\\
S^{-1}\, \pi^{(2)}\, S &=& \sum_{a,a'}
E^{a a}\otimes E^{a'a'}+\sum_{\beta,\beta'}
E^{\beta \beta}\otimes E^{\beta'\beta'}
+\sum_{a} \sum_{\beta}  E^{a a}\otimes E^{\beta \beta}\label{diagp2}
\eea

To use the  fusion formulae   for
the Hubbard  models one  doubles every space to 
unprimed and primed copies. For instance $R_{14}(\la,\mu-\rho)$
in (\ref{fus4}) is replaced by $R_{11'44'}(\la,\mu-\rho)$ 
or  $r_{11'44'}(\la,\mu-\rho)$, obtained
from (\ref{rh}) or (\ref{rrh}), respectively. 
The calculations involved in (\ref{fus1}) and (\ref{fus4}) 
are straightforward and can be
carried out using the explicit expressions
(\ref{proj1},\ref{proj2},\ref{ss},\ref{ss1}). 
Similarly, the quadratic Hamiltonian density 
can eventually be  obtained from the right-hand side of 
(\ref{hamden}) or (\ref{hamden2}),
or directly once  (\ref{fus4}) is calculated. 
The explicit expressions in terms of $E$-matrices are however unwieldy, complicated, and unenlightening unless used for specific applications
such as  writing down the Hamiltonian in terms of 
higher-spin  $sl(2)$ generators, or for an explicit diagonalization. 

It is easy to verify 
that the relation between the fused matrices  based on $R$ and $r$ is 
a non-diagonal similarity transformation:
\bea
R^{(i)}_{<12><34>}(\la_1,\la_2)&=&
S_{12}^{-1} S_{1'2'}^{-1} S_{34}^{-1}  S_{3'4'}^{-1}\,
I_{11'}(h_1)I_{22'}(-h_1) I_{33'}(h_2)I_{44'}(-h_2)\nonumber\\
&\times & S_{12} S_{1'2'} S_{34} S_{3'4'}\, 
r^{(i)}_{<12><34>}(\la_1,\la_2)\,
S_{12}^{-1} S_{1'2'}^{-1} S_{34}^{-1}  S_{3'4'}^{-1}\nonumber\\
&\times& I_{11'}(-h_1)I_{22'}(h_1) I_{33'}(-h_2)I_{44'}(h_2)\,
S_{12} S_{1'2'} S_{34} S_{3'4'}\label{gagh}
\eea
To unravel the coupling structure of the models just obtained 
from fusion, we can look for Lax matrices as in (\ref{rtol}). 
The following relations for (\ref{ffr}) are  easily derived:
\bea
& & R_{12}(\la\pm\frac{\pi}{2})C_1 = -C_1 R_{12}(\la\mp\frac{\pi}{2})\\
& & R_{12}(\la\pm\frac{\pi}{2})C_1 = - R_{12}(\pm\frac{\pi}{2}-\la) C_2
\eea
(Such relations clearly have fermionic counterparts \cite{pz}.)
One can then obtain the fused matrices (\ref{fus4}) (for (\ref{rh})) at
some particular points:
\bea
R^{(i)}_{<12><34>}(\la,0)&=&\frac{1}{\cosh^2 h}\, S_{12}^{-1} 
S_{1'2'}^{-1}I_{11'}(-h)I_{22'}(h) S_{12}S_{1'2'}
\, R^{(i)}_{<12><34>}(\la)\nonumber\\
&\times &  R^{(i)}_{<1'2'><3'4'>}(\la)
S_{12}^{-1}S_{1'2'}^{-1}I_{11'}(h)I_{22'}(-h)S_{12}S_{1'2'}\label{sp1}
\eea
and
\bea
R^{(i)}_{<12><34>}(\la,\frac{\pi}{2})&=&\frac{1}{\cosh^2 h}\, S_{12}^{-1} 
S_{1'2'}^{-1} I_{11'}(h)I_{22'}(-h)S_{12}S_{1'2'}
\, R^{(i)}_{<12><34>}(\la-\frac{\pi}{2})\nonumber\\
&\times & 
R^{(i)}_{<1'2'><3'4'>}(\la-\frac{\pi}{2})
S_{12}^{-1}S_{1'2'}^{-1}I_{11'}(-h)I_{22'}(h)S_{12}S_{1'2'}\label{sp2}
\eea
These expressions correspond to the decoupled product of two
copies of multi-flavor spin-$1$ ($i=2$), or spin-$0$ ($i=1$)
models at their free-fermion point $\gamma=\pi/2$. Indeed, contrary
to what happens in (\ref{rtol}), 
the product of $SIS$'s on the right is the inverse of that on the right,
and as such  the $I$-matrices implement an innocuous gauge transformation
rather than a coupling. (Another explanation of the non-coupling nature of
the $I$'s is found in the following paragraph.) 
This  negative result can be understood with 
hindsight.  A simple coupling through  the $I$-matrices would have been
naive because the conjugation operator $C$ has very special properties
with respect to the $R$-matrices and is peculiar to the spin-$\half$
representation. 

The points $\mu=0,\pm\pi/2$, and  similarly $\la=0,\pm\pi/2$, 
are then decoupling points. For a generic pair $(\la,\mu)$ 
there is no decoupling, and (\ref{fus1},\ref{fus4}) applied
to the Hubbard models yield the multi-flavor spin-$(0\times\half)$,
spin-$(1\times\half)$, spin-$(0\times 0)$ and  
spin-$(1\times 1)$ Hubbard models. 
The structure of their corresponding  matrices has a simple 
interpretation.  As emphasized in section \ref{hubpri}, 
it is the special  linear combination of $RR$ and $RCRC$ appearing 
in (\ref{rrh}) which is  characteristic of the Hubbard models.  
Looking back at (\ref{fus4}) and expanding  the product with $R$
replaced by $r$, one finds a  special linear combination of sixteen terms, 
each a product of $RR$ and $RCRC$. The sixteen terms precisely
account for all possible combinations one could have found natural to consider. 
Both $R$ and $RC$  satisfy Yang-Baxter equations \cite{sh3,ff},
and are therefore candidates for  linear combinations. Fusion nicely retains 
this structure. It also shows the preponderant role played by the 
underlying $sl(2)$ structure of the XX models which can be repeatedly 
fused to reach any spin and therefore any spin-$(s\times s')$ Hubbard model.

\subsection{Non-locality at $q^2=-1$}\label{curi}

An important issue is whether a given $R$-matrix is regular.
Within the QISM, integrability ensures the existence of a large number
of commuting  quantities. The Yang-Baxter equation
implies that the transfer matrices for $N$ sites,
$\tau(\la,\mu)= {\rm Tr}_0 \left[ R_{0N}(\la,\mu)\cdots R_{01}(\la,\mu)
\right]$,
mutually commute at arbitrary values of either of the spectral parameters
(the other one remaining fixed). The matrix $\tau(\la,\mu)$
is therefore the generator of mutually commuting quantities. 
Such quantities are generically  non-local. For periodic 
boundary conditions, local, commuting spin-chain Hamiltonians  can however
be defined by taking the derivatives of the logarithm of the transfer
matrix at a point where the $R$-matrix is regular. 
The transfer matrix at such a point is proportional to the unit-shift
operator on the chain. Its inverse, in the logarithmic derivatives,
`cancels' most of the operators in the numerators at the sites where
no derivative has been taken. In the following cases regularity does
not hold. Unitarity by itself is not enough to ensure that the
transfer matrix is invertible, and each case should be considered separately. 

Before turning to the fused Hubbard models, consider 
the spin-$1$ matrix which can be obtained  by fusion from the XX models
(\ref{ffr}), and which appears in (\ref{sp1},\ref{sp2}) for $i=2$. 
The following gauge transformation turns the asymmetric $m=2$ XXC
$R$-matrix  of \cite{zmfuse} into a  symmetric matrix $R^{(s)}$:
\beq 
R^{(s)}(\la)=(\bI\otimes A(\la))\, R(\la)\, (\bI\otimes A(-\la))
\eeq
where $A(\la)=\sum_{\af_1}E^{\af_1\af_1}e^{ic_1\la}+
\sum_{\af_2}E^{\af_2\af_2}e^{ic_2\la}$ and  $c_2-c_1=1$.
At $\gamma=\pi/2$,
\beq
R^{(s)}(\la)=\tilde{P}^{(1)}\sin\gamma +
\tilde{P}^{(2)}\sin(\gamma+\la)+\tilde{P}^{(3)}\sin\la
\eeq
reduces to (\ref{ffr}). The net effect on fusing the symmetric version,
for any value of $q$,
is to remove (before $q^2\rightarrow -1$) all factors of 
$y^{\pm 1}$ and $q^{\pm 1}$ from (38) in \cite{zmfuse}.
The symmetric spin-$1$ matrix is reproduced in the appendix. 
For $\gamma=\pi/2$ the two simple zeroes of $f(\la)=\sin(\gamma+\la)
\sin(\gamma-\la)$   become a double zero at $\la=\pi/2$. 
Setting $\gamma=\pi/2$ in (\ref{rapp}) allows to cancel out a factor
of $\sin\la$. The resulting matrix satisfies the Yang-Baxter equation,
is  unitary but not regular at any value of  $\la$. 
However one can  still define  local Hamiltonians  through a limiting
procedure from generic values of $\gamma$. One drops the  prefactor in the 
left-hand side of (\ref{rapp}), and calculates the local conserved 
quantities  with this renormalized $R$-matrix:
\beq
H_{p+1}=(\sin\gamma\sin 2\gamma)^p\frac{d^p}{d\la^p}\log\left({\rm Tr}_0 
\left[ R_{0N}(\la)\cdots R_{01}(\la)\right]\right)_{|\la=0}\;\;,\;\;\; p\geq 0 
\eeq
These   commuting
local Hamiltonians are finite and non-trivial 
as $\gamma\rightarrow \pi/2$. 
(The factor $(\sin\gamma\sin 2\gamma)^p$
may cancel some contributions but leaves the main ones.)
Thus despite the lack of  a regular point it is possible to define
local conserved quantities.

The function $f(\la_1,\la_2)$ in (\ref{hunit}) 
has a double zero,  at $\la_1-\la_2=\pm\pi/2$,  which is inherited from
the XX models forming the Hubbard matrix. 
The special cases (\ref{sp1},\ref{sp2})  are non-regular matrices 
and indicate that the fused matrix (\ref{fus4}) ($i=2$)
for the Hubbard models, is probably  not regular for
all values of $\la=\mu$. (For the spin-$(0\times 0)$ case, $i=1$,
dimensionality considerations imply regularity, just
as for the XX models.) A definite proof of non-regularity 
for generic values of the spectral parameters would be welcome.  
There is no known `quantum' deformation of the generalized
Hubbard models to invoke a  first order zero and take the limit. 
Although the absence of local quantities would be surprising, 
it is  not clear whether they exist and how they can be
calculated for all the Hubbard models
corresponding to coupled spin-$(s\times s')$ XX models, with 
$s\; {\rm or}\; s'\geq 1$. 

Finally, consider the point $q^2=-1$ ($\gamma=\pm\pi/2$) for all the 
models considered in \cite{zmfuse}. The function
$f(\la)$ in the unitarity equation picks up a double zero at $\la=\pi/2$.  
So one can expect the loss of the regularity property 
for all the matrices $R^{(i)}_{<12><34>}(\la)$ obtained from 
the $R$-matrices of the defining representations 
of the multiplicity $A_m$ models. 
This was seen  above explicitly 
for the spin-$1$ $A_1$ model. 
Further fusions will propagate this non-regularity to all the higher 
representations, with the  combined appearance of higher 
harmonics of $\gamma$, that is of higher roots of unity for $q$.  
Local conserved quantities should however still be obtainable
through the limiting method described above.  
That the fourth roots of unity, and more generally $n^{\rm th}$ roots of unity,
play a specific role for the single flavor models ($n_i=1,\; i=1,\cdots,m+1$)
is not surprising. The representation theory of 
the quantum algebra $\CU_q(sl(m+1))$ is in one-to-one correspondence
with the one for the undeformed algebra when  $q$ is
not a root of unity. This 
is not the case for roots of unity: the representation theory becomes 
richer and more complicated. The  defining representations are
however undeformed for all values of $q$. 

\section{Conclusion}

The fusion method was shown to generalize naturally to
non-additive solutions of the Yang-Baxter equation. Expressions
for the fused matrices,  the regularity and unitarity equations and
the quadratic Hamiltonians  were obtained.  
The issue of non-simple zeroes was raised 
and connected to a possible lack of regularity of the fused matrices. 
This raised the issue of  existence of a set of {\it local}
commuting quantities.  
The generalized Hubbard  models  were then shown to allow fusion
for all spin-$(s\times s')$ representations,  and
compact expressions were  obtained for the $R$-matrices corresponding
to mixed spin-$0,\half,1$ multi-flavor representations. 
Local Hamiltonians are believed to exist but a definite proof 
and calculation method are lacking. 

The fused  Hubbard  models  inherit the symmetries of the two
coupled  multi-flavor spin-$(s\times s')$ copies. 
These symmetries, and the fusion equations between the various 
transfer matrices can be used to  diagonalize them 
through the  algebraic Bethe Ansatz. (The lack of regularity 
should not pose a problem.) 

The connection, noted in section \ref{hubpri} for  the Hubbard models,
between the $L$ and $R$ matrices
and the corresponding $RLL$ and $RRR$ Yang-Baxter equations
could serve as a naturalness test for the choice of an $R$-matrix. 
This could be particularly relevant in the Bariev model for 
which more than one $R$-matrix is known to exist \cite{bariev,hqz,sw}.
More generally one can ask the following question.  Provided 
the dimensions match and given a Lax operator associated to a
non-additive $R$-matrix, can $L$ be obtained as a  limiting case
of the original $R$ or some other one with the $RLL$ relation satisfied ? 
Another general test could be  the existence of  projector points. 
Fusion should also be applicable to the Bariev and Chiral Potts models.  

Finally,  a non-additive  matrix for the Bariev model was recently obtained
by twisting a quantum group $R$-matrix and taking a singular
limit \cite{links}. It is however not clear whether such a method can be
made  to work for the Hubbard models. 
\\

\noindent{\bf Acknowledgement:} I would like to thank P. Fendley and
I. Horvath for a critical reading of the manuscript. 
\\

\noindent{\Large \bf Appendix: The symmetric multi-flavor spin-$1$ matrix}
\\

\noindent The spin-$1$ matrix discussed in section \ref{curi} is given by:
\bea
\frac{\sin(2\gamma-\la)}{\sin(\la+\gamma)}&\times & 
R_{<12><34>}^{(2)(s)}(\la) =\label{rapp}\\
& & +\sin(\la+\gamma) \sin(\la+2\gamma)
\sum_{a,b,c,d} E^{ab}\otimes 
E^{cd}\otimes E^{ba}\otimes 
E^{dc}\nonumber\\
& & +\sin(\la+\gamma) \sin(\la+2\gamma)
\sum_{\af,\be,\gamma,\delta} E^{\af\beta}\otimes 
E^{\gamma\delta}\otimes E^{\be\af}\otimes E^{\delta\gamma}\nonumber\\
& & + \sin(2\gamma)\sin(\la+\ga)
\sum_{a}\sum_{\af,\be,\gamma} E^{\af a}\otimes 
E^{\be\gamma}\otimes E^{a\af}\otimes E^{\gamma\be}\nonumber\\
& & + \sin(2\gamma)\sin(\la+\ga)
\sum_{a,b,c}\sum_{\af} E^{ab}\otimes 
E^{c\af}\otimes E^{ba}\otimes E^{\af c}\nonumber\\
& & + \sin(2\gamma)\sin(\la+\ga)
\sum_{a,b,c}\sum_{\af} E^{ab}\otimes 
E^{\af c}\otimes E^{ba}\otimes E^{c \af}\nonumber\\
& & + \sin(2\gamma)\sin(\la+\ga)
\sum_{a}\sum_{\af,\be,\gamma} E^{a \af}\otimes 
E^{\be\gamma}\otimes E^{\af a}\otimes E^{\gamma\be}\nonumber\\
& & +(\sin\gamma \sin(2\gamma)+\sin\la\sin(\la+\ga))
\sum_{a,b}\sum_{\af,\be} E^{ab}\otimes 
E^{\af\be}\otimes E^{ba}\otimes E^{\be\af}\nonumber\\
& & + \sin\gamma\sin(2\gamma)
\sum_{a,b}\sum_{\af,\be} E^{\af a}\otimes 
E^{\be b}\otimes E^{a \af}\otimes E^{b \be}\nonumber\\
& & + \sin\gamma\sin(2\gamma)
\sum_{a,b}\sum_{\af,\be} E^{a\af}\otimes 
E^{b\be}\otimes E^{\af a}\otimes E^{\be b}\nonumber\\
& & +2 x^{-1} \cos\gamma \sin(2\gamma)\sin\la\,
\sum_{a,b}\sum_{\af,\be} E^{\af a}\otimes 
E^{\be\af}\otimes E^{ab}\otimes E^{b\be}\nonumber\\
& & +x  \sin\gamma\sin\la\,
\sum_{a,b}\sum_{\af,\be} E^{ab}\otimes 
E^{\af a}\otimes E^{b \be}\otimes E^{\be \af}\nonumber\\
& & +x^{-2} \sin\la\sin(\la+\gamma)
\sum_{a,b,c}\sum_{\af} E^{ab}\otimes 
E^{\af\af}\otimes E^{bc}\otimes E^{ca}\nonumber\\
& & +x^{-2} \sin\la\sin(\la+\gamma)
\sum_{a}\sum_{\af,\be,\gamma} E^{\af\be}\otimes 
E^{\gamma\af}\otimes E^{aa}\otimes E^{\be\gamma}\nonumber\\
& & +x^2 \sin\la\sin(\la+\gamma)
\sum_{a,b,c}\sum_{\af} E^{ab}\otimes 
E^{ca}\otimes E^{bc}\otimes E^{\af\af}\nonumber\\
& & +x^2 \sin\la\sin(\la+\gamma)
\sum_{a}\sum_{\af,\be,\gamma} E^{aa}\otimes 
E^{\af\be}\otimes E^{\be\gamma}\otimes E^{\gamma\af}\nonumber\\
& & +x^{-3}  \sin\gamma\sin\la\,
\sum_{a,b}\sum_{\af,\be} E^{a\af}\otimes 
E^{\be\be}\otimes E^{bb}\otimes E^{\af a}\nonumber\\
& & +2 x^3  \cos\gamma \sin(2\gamma)\sin\la\,
\sum_{a,b}\sum_{\af,\be} E^{aa}\otimes 
E^{b\af}\otimes E^{\af b}\otimes E^{\be\be}\nonumber\\
& & +x^{-4} \sin(\la-\gamma)\sin\la\,
\sum_{a,b}\sum_{\af,\be} E^{\af\af}\otimes 
E^{\be\be}\otimes E^{aa}\otimes E^{bb}\nonumber\\
& & +x^4 \sin(\la-\gamma)\sin\la\,
\sum_{a,b}\sum_{\af,\be} E^{aa}\otimes 
E^{bb}\otimes E^{\af\af}\otimes E^{\be\be}\nonumber
\eea
This matrix  is regular for $\gamma
\neq\frac{\pi}{2}+k\pi$ ($k\in\bZ$),  
and unitary for arbitrary $\gamma$:  
\bea
& & R^{(2)(s)}_{<12><34>}(0) =  \sin^2\gamma\; \CP_{13} \CP_{24}\, 
\pi^{(d)}_{12}\pi^{(d)}_{34}\\
& & R^{(2)(s)}_{<12><34>}(\la)  R^{(2)(s)}_{<34><12>}(-\la) = \sin^2(\gamma+\la)\sin^2(\gamma-\la)\,
\pi^{(d)}_{12} \pi^{(d)}_{34}  
\eea
where $\pi^{(d)}$ is equal to the right-hand side of (\ref{diagp2}).
(It is necessary to let $x\longrightarrow -x$ before using (\ref{rapp})
in (\ref{sp1},\ref{sp2}).)

\end{document}